\documentclass[%
 reprint,
 amsmath,amssymb,
 aps,
]{revtex4-2}

\usepackage{graphicx}% Include figure files
\usepackage{dcolumn}% Align table columns on decimal point
\usepackage{bm}% bold math
\begin{document}

\preprint{APS/123-QED}

\title{Demographics of co-aging complex systems: from sickly worms to chess engines}% Force line breaks with \\

\author{Cagatay Eskin}
\email{ceskin@nd.edu}
\author{Dervis Can Vural}%
 \email{Corresponding author: dvural@nd.edu}
\affiliation{%
 Department of Physics and Astronomy, University of Notre Dame, Notre Dame, Indiana 46556, USA\\
}%

\date{\today}% It is always \today, today,

\begin{abstract}
Aging, as defined in terms of the slope of the probability of death versus time (hazard curve), is a generic phenomenon observed in nearly all complex systems. Theoretical models of aging predict hazard curves that monotonically increase in time, in discrepancy with the peculiar ups and downs observed in empirically. Here we introduce the concept of co-aging, where the demographic trajectories of multiple cohorts couple together, and show that co-aging dynamics can account for the anomalous hazard curves exhibited by some species. In our model, multiple interdependency networks inflict damage on one other proportional to their number of functional nodes. We then fit our model predictions to three datasets describing (1) co-aging worm-pathogen populations (2) competing tree species. Lastly, we gather the mortality statistics of (3) machine-against-machine chess games to demonstrate that co-aging dynamics is not exclusive to biological systems.
\end{abstract}

\maketitle

%\tableofcontents

%\section{\label{sec:level1}First-level heading:\protect\\ The line
%break was forced \lowercase{via} \textbackslash\textbackslash}

\section{\label{sec:Introduction}Introduction}
Typically, simple systems consisting of few constituents, such as radioactive nuclei or a sledgehammer, will fall apart with constant probability per unit time. In contrast, the failure probability of a complex system such as a human or a jet engine tends to increase by many folds over its life span. The positive slope of failure probability versus time (called the ``hazard curve''), is the demographic definition of aging. Radioactive nuclei do not age, whereas jet engines do.

Evolutionary theories of aging offer a plausible explanation for this positive slope: Since an organism risks experiencing an external hazard at any point in its life, it is a better evolutionary strategy to reproduce early on rather than later in life. In such early-reproducing lineages, natural selection cannot eliminate late-acting deleterious traits, which manifest as aging (mutation accumulation theory \cite{Medawar1952,Williams_1957,Hamilton_1966,medawar1946old}). Some late-acting deleterious genes might even happen to enhance early-life success, thereby get positively selected (antagonistic pleitropy theory  \cite{Williams_1957,Rose1991,Kirkwood_2002}).

In \cite{vural2014aging} it was argued that aging is a generic phenomenon characteristic of systems consisting of a large number of interdependent components. Networks of interdependence are fragile: If one component malfunctions, so will others that crucially depend on it. The failure statistics of such networks was shown to accurately describe the demographic trajectories of biological species as well as complex mechanical devices \cite{vural2014aging}. The interdependence network picture does not negate the evolutionary arguments outlined above, but works in tandem; and has since been fruitful in progressing our empirical and theoretical understanding of aging populations \cite{vural2014aging, Acun_2017,suma2018interdependence,uppal2020tissue,Farrell_2016,Taneja_2016,Mitnitski_2017,Farrell_2018,Farrell_2020,Farrell_2021} (also see the appendix of \cite{stroustrup2016temporal} for a direct test of the model).

\begin{figure}%[tbhp]
\centering
\includegraphics[width=\linewidth]{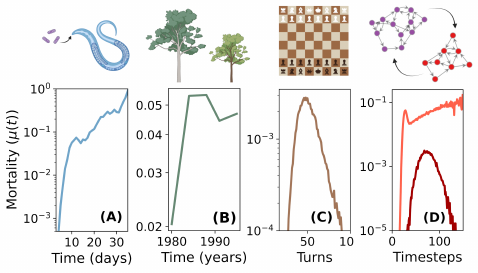}
\caption{{\bf The co-aging systems we study, alongside their typical hazard curves}. (A) \textit{C. elegans} colonized by pathogenic \textit{E. coli}. The worm's mortality curve exhibits a distinct early life ``kink''. (B) \textit{Acer macrophyllum}, the Bigleaf maple, during the thinning phase of their lifecycle, exhibit a similar non-monotonic mortality pattern, as they experience heightened competition for resources. (C) Chess engines. Chess defense networks inflict damage on each other similar to biological antagonists, thereby exhibiting similar non-monotic mortality curves. (D) Our theoretical model where two interdependency networks inflict damage on each other proportional to the number of their functional nodes. The two curves correspond to two different parameter sets.}
\label{fig:figure1_systems}
\end{figure}

Evolutionary and interdependency network models both predict hazard curves that monotonically increase in time. However, empirically, hazard curves with peculiar ups and downs have been observed, in apparent contradiction with theory \cite{Jones_2014, Vaupel1998}. Some organisms have higher probability of death when younger, while others start anti-aging midlife, only to continue aging later. How should we interpret and predict such features?

We hypothesize here that a non-monotonic hazard curve signals a coupling between the aging processes of multiple species. For example, some tree species in early successional forests exhibit non-monotonic hazard curves during the ``thinning'' phase of stand development \cite{Lutz_2006}. The primary factor contributing to mortality at this stage is suppression, arising from intense competition for vital resources such as nutrients, water, and sunlight \cite{McMurtrie_1981,Peet_1987,Franklin_1987}. Another example is worms experiencing an increase in early-life mortality due to bacterial infections. These infections not only manifest as faster aging and shorter lifespan, but also as qualitatively distinct kinks or bumps in the worm's hazard curve \cite{ZhaoYuan2017Tfod, Baeriswyl_2009}. 

In this paper we introduce the concept of co-aging, and offer a theoretical model to interpret and predict the demographic data of co-aging populations. In our model, two fragile interdependency networks assault each other and influence each other's failure statistics. Then using this model, we analyze three empiric data sets: two from the available literature, and one gathered by us. (Fig.\ref{fig:figure1_systems}).

The first data set \cite{ZhaoYuan2017Tfod}, measures the hazard curves of \emph{C. elegans} whose digestive tracks are colonized by pathogenic \emph{E. coli}. The immune system of the worm inflicts damage on the colony, and vice versa, leading to an anomalous hazard curve. The second data set \cite{Lutz_2006}, measures the demographic trajectories of three different species of trees from early successional forests. This data spans the ``thinning'' stage which is known as the dynamic period of stand development in trees. During the thinning stage, one of the most prominent drivers of the mortality is inter- and intraspecific competition. We argue here that the unusual hazard curves observed in many tree species is due to co-aging, i.e. individuals indirectly damaging each other by cutting off access to vital resources.

One of the appeals of our interdependecy picture over its evolutionary counterparts is its applicability beyond biology. Just as aging can be observed in non-biological complex machines \cite{vural2014aging}, we argue here, so can co-aging. To demonstrate this, as a third study case, we gather the failure statistics of machine-vs-machine chess players, where, much like the worm-pathogen and tree-tree systems, the players continuously penetrate each others' network of defenses, and exhibit similar co-aging signals in their demographic trajectories.

We shall see that worms and pathogens, competing trees, and chess battles all have similar hazard curves, and can all be accurately fit to our co-aging interdependency network theory presented here. 

Furthermore, our model can account for the difference in hazard curves of infected (co-aging with bacteria) and antibiotic-fortified (non-co-aging) worms by simply turning down a single parameter (that quantifies the antagonist-inflicted damage). It can, by also changing the same single parameter, account for the difference in hazard curves of chess engines with different thinking depths: Higher thinking depth enables larger impact per piece, hence proximally quantifies the antagonist-inflicted damage.

\section{Network Model of Co-Aging}
Our model consists of two complex networks, each comprising of nodes and directional edges, which we interpret as functional constituents (such as genes or cells) and the interdependency between them. %In this study, we have built the network system using a framework from previous research on aging \cite{vural2014aging}. However, we have made significant modifications to the framework to account for co-aging. For this study, we have used a directed random network topology and followed the standard procedures outlined in \cite{vural2014aging} to construct the network.

{\bf Initialization of networks (evolutionary time scales).}  We begin by creating a large cohort of network pairs representing two adversarial species, $A$ and $B$, with number of nodes $N_A$ and $N_B$. All members of a species are assumed to have identical network structure. For large networks, randomizing network topology between individuals yields identical results, with insignificant additional noise (see supplemental material). 

Networks are built by adding one node at a time. To avoid coincidental occurrences of disconnected sub-networks, each newly added node connects inwards randomly to one existing node and outwards to another.

This random growth of interdependency represents a ``constructive neutral'' evolutionary process, first (verbally) hypothesized by \cite{stoltzfus1999possibility}. It was shown earlier \cite{vural2014aging} that a non-neutral growth (where the connection probability is non-uniform) leads to similar outcomes in mortality statistics; so we do not explore here the effect of topology further.

{\bf Aging of networks (individual time scales).} After growing the networks, as if by constructive neutral evolution \cite{stoltzfus1999possibility}, we probe their fragility; alone (aging) and together (co-aging). During the course of an aging simulation, every node assumes one of two states: functional, or dysfunctional. 

%Each node is assigned one of three values: $x_i \in \{f\equiv1, n_s\equiv0, n_c\equiv0\}$. $f\equiv1$ labels a functional node, while states $n_s\equiv0$ and $n_c\equiv0$ correspond to nonfunctional nodes due to intrinsic aging and co-aging, respectively. At birth, each type of networks will have a certain percentage of nonfunctional nodes, denoted as $f_A(0)$ and $f_B(0)$. Depending on the percentage of prenatal damage to the complex system, $f_{A,B}(0)$ some of the node states are initially marked as $n_s\equiv0$, corresponding to nonfunctional nodes due to intrinsic aging, while all other nodes are functional, namely having a state of $f\equiv1$. We measure the networks' vitality as $\phi_A(t)=\sum_i x_i^A(t)/N_A$ and $\phi_B(t)=\sum_j x_j^B(t)/N_B$, where $\vec{\psi_A}(t)=\{x_1^A(t), x_2^A(t),..., x_{N_A}^A(t)\}$ and $\vec{\psi_B}(t)=\{x_1^B(t), x_2^B(t),..., x_{N_B}^B(t)\}$ are defined as the organism state.

We age networks as follows: At each time step, some of the nodes in $A$ will be marked to malfunction. This is done in two rounds. In the first round, each node is marked with probability $d_A$. This is the steady damage rate experienced by the organism in the absence of its adversary. We then iteratively ``propagate'' this damage throughout the network, by also marking the nodes that lost more than half of their dependees.

In the second round, additional nodes are marked for failure with a probability $\alpha_{BA}(t)$ proportional to the fraction of functional nodes $f_B$, in the opponent network $B$, namely, $\alpha_{BA}(t)=C_A f_B/N_B$, where the ``co-aging parameter'' $C_A$ quantifies the aging impact of $B$ on $A$. This damage is also propagated in a similar fashion to round one.

After each round, we also allow for the damaged node to revert to its functional state with probability $r_A$. This is the repair rate.

At the end of these two rounds, we update the labels of the marked nodes. If a node was marked for failure during the first round or due to a majority of its dependees malfunctioning during the first round (even though it might have failed in a second round), we register its cause of death as ``intrinsic damage'', and otherwise, as ``co-aging damage''. For the dysfunctional nodes whose dependees were damaged half and half by intrinsic and co-aging damage: if the node was marked for failure during a first round, we register the cause of death as intrinsic aging. If marked during a second round, co-aging.

%Eskin: I tried to fix this paragrah by stating that we initiane repair after both intrinsic aging and co-aging damages. This may seem like we are doing to repairs in one step but as both intrinsic aging and co-aging damages present at one time step, for each one of those there will be a repair. So, we end up having two damage and two repair steps at a time. Because of that units stay consistent. Maybe we can add this repair step in to above paragraph.

Then, $A$'s antagonist $B$ also undergoes damage and repair in identical fashion, with analogous damage rates $d_B$, $\alpha_{AB}(t)$, 
and repair rate $r_B$; and one time step is complete. Then, we move on to the next time step and repeat this process until both of the networks die. 

We proclaim a network dead if the fraction of its functional nodes go below 1\%. Since the overwhelming majority of networks collapse suddenly once the functional nodes fall below 40\%-60\% (see Fig.\ref{fig:figure2_mortalities_and_vitalities} insets), our outcomes do not sensitively depend on the value of this threshold (see supplemental material). 

Once both networks die, we record their age of death and cause of death, and move on to another pair and age them the same way.  %we define something "dead" if fraction of alive nodes drop below 0.1. Why not zero? how sensitive is the shape of the mortality curves to the value of this chosen threshold?

We keep track of the cause of malfunction of the nodes so that we can tell the cause of death of each network. This will later enable us to fit the data of \cite{ZhaoYuan2017Tfod} who were able to resolve between worms dying of old age (intrinsic damage) versus infection (co-aging damage). This way, our theory will be able to fit not only cumulative hazard curves, but also, simultaneously, the cause-specific hazard curves (Fig.\ref{fig: figure_elegans}). Furthermore, we will even be able to fit the antibiotic treated worms (which do not co-age) and non-treated worms (which do co-age) using the same parameter values, except for simply turning down the bacterial impact, $C_{B}$.

The number of functional, and two types of dysfunctional nodes (due to intrinsic aging and co-aging) at time $t$ are denoted by $f(t)$, $n_s(t)$, $n_c(t)$. When a network collapses, we record whether its failure was mostly due to intrinsic damage (if $n_s>n_c$) or co-aging damage (if $n_s<n_c$), which defines in our model, the cause of death. 

Once the entire cohort dies, the hazard curves (probability of death) $\mu(t)=[S(t)-S(t+1)]/S(t)$ are obtained in terms of the number of survivors $S(t)$ at $t$. The cause-specific hazard curves are defined similarly,
$\mu_s(t)= s(t)/S(t), 
\mu_c(t)= c(t)/S(t)$
in terms of the number of individuals that die of intrinsic damage $s(t)$ and co-aging damage $c(t)$.

\section{Results}
\begin{figure}%[tbhp]
\centering
\includegraphics[width=\linewidth]{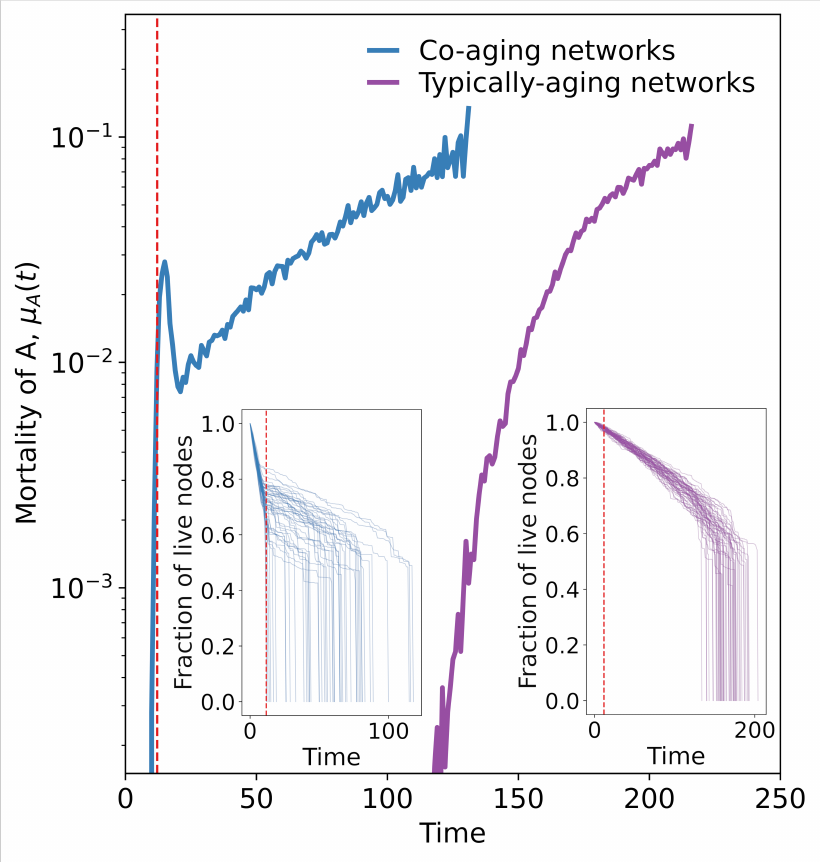}
\caption{{\bf The hazard curves of co-aging (left) and typically aging (right) complex networks}. A kink or bump is the hallmark of co-aging, and occurs near the mean lifespan the antagonist (vertical dashed line). The insets show the decline of function of fifty individual networks, where co-aging manifests as a sudden reduction of slope coinciding, again, with the mean lifespan of the antagonist.  The right curve and right inset describe a network aging solo, which lacks these peculiar features. Co-aging parameters are  $C_A=0.015$ (left) vs. $C_A=0$ (right), while other parameters (see supplemental material for values) are kept constant.} 
\label{fig:figure2_mortalities_and_vitalities}
\end{figure}

{\bf Overview.} In our simulations, the fraction of nodes alive $f(t)$, declines gradually, followed by a sudden collapse (Fig.\ref{fig:figure2_mortalities_and_vitalities} insets). When we compare $f(t)$ for species aging in solitude versus co-aging with an antagonist, we see that the latter exhibits a sharp change in slope coinciding with the average life-span of its antagonist (Fig.\ref{fig:figure2_mortalities_and_vitalities}, vertical dashed line).

Fig.\ref{fig:figure2_mortalities_and_vitalities} depicts the hazard curves for species $A$ with (left) and without (right) its co-aging antagonist. The insets show the functional decline of $50$ randomly chosen $A$ individuals. The vertical dashed line marks $\tau_B$, the mean lifespan of $B$. In the left inset, nearly all curves exhibit a sharp change in slope after the expected lifespan of the antagonist, as co-aging damage diminishes after this time.

In Fig.\ref{fig:figure2_mortalities_and_vitalities}, the difference between the two hazard curves is solely due to the value of the co-aging parameter $C_A$, which causes the hazard curve to have a ``bump'' around $\tau_B$, the mean life span of the antagonist. Here $\mu(t)$ increases from birth till $\tau_B$ due to the combined effect of the intrinsic and coaging damage; starts to decrease as the antagonists perish (due to intrinsic and co-aging damage of their own), followed by, again, a late-life increase in mortality due to the system's intrinsic damage. 

The hazard curve on the right was run with the same parameters except for $C_A$ set to zero. With this, the bump vanishes, and we recover aging trajectories observed more typically, as in \cite{vural2014aging}. 

\begin{figure}%[tbhp]
\centering
\includegraphics[width=1\linewidth]{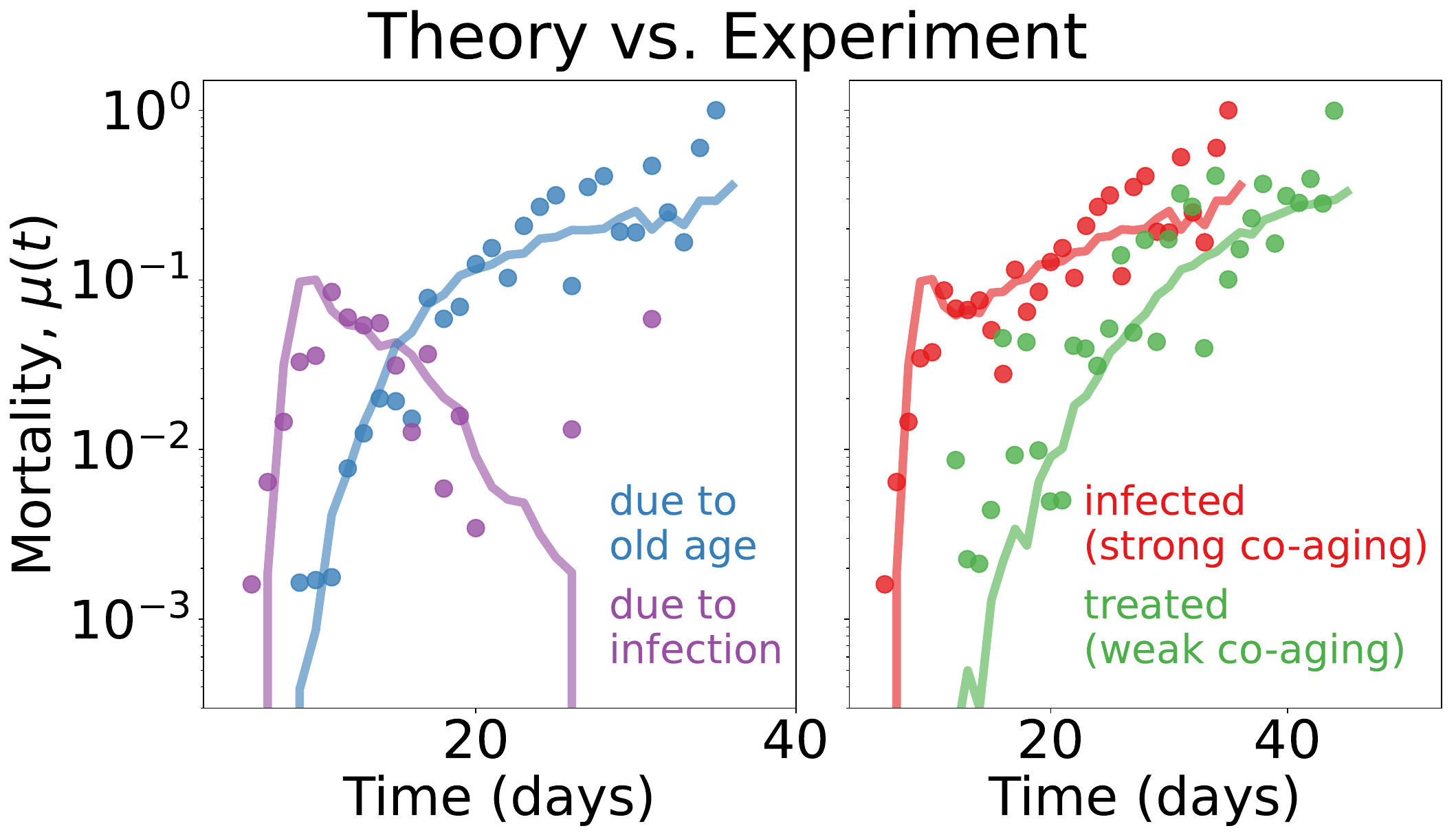}
\caption{{\bf Co-aging of \emph{C. elegans} and its pathogen \emph{E. coli} (dots), fit to co-aging networks (curves).} Left: The decomposition of cumulative hazard curves of the worms into cause-specific hazard curves. As experimental measurements can go beyond counting total deaths and resolve cause of deaths \cite{ZhaoYuan2017Tfod}, we have also labeled the cause of deaths of our simulated networks, which agree well with experiment. Right: Cumulative mortality (sum of cause-specific hazards in the left panel) for infected vs. antibiotic-treated worm populations. Modifying the theoretical red curve (that fits infected worms) to yield the green one (that fits antibiotic treated worms) was achieved by turning down a single parameter, the co-aging constant, which quantifies the impact of an antagonist. All other parameters (see supplemental material) were kept constant.}
\label{fig: figure_elegans}
\end{figure}

{\bf Disentangling the two types of aging in worms.} Recently, \cite{ZhaoYuan2017Tfod} measured the hazard curves of a \emph{C. elegans} population coexisting with their pathogen \emph{E. coli} and those that were treated with antibiotics so that they age intrinsically. They also recorded the cause of death of the worms through necropsy analysis, by observing severe swelling of the posterior pharyngeal bulb in individuals who died earlier in life due to infection, and atrophy of the same region in those who died later due to other causes (in our interpretation, due to intrinsic damage). During early life, \textit{C. elegans} exhibits rapid pharyngeal pumping which can damage the pharyngeal cuticle, making individuals more susceptible to \textit{E. coli} invasion. However, due to heterogeneity in the population, some individuals are able to resist this invasion, while others get colonized and die. This way, \cite{ZhaoYuan2017Tfod} was able to report cause-specific hazard curves as well as cumulative ones. For the details of population heterogeneity in the context of \textit{C. Elegans}, see \cite{Cunff_2013, Cunff_2014}

In Fig.\ref{fig: figure_elegans} we fit all empirical measurements of \cite{ZhaoYuan2017Tfod} using our theory. In the left panel, we have cause-specific hazard curves, i.e. counting the individuals that die due to co-aging versus intrinsic aging. On the right panel, we fit the total mortality curves (summing co-aging and intrinsic aging together) in the absence and presence of antibiotics. We should emphasize that we fit all data using a single combination of parameters, except for reducing the parameter that quantifies the bacteria-inflicted damage ($C_A=0.023$ to $C_A=0.013$) to account for the presence of antibiotics in some of the experiments.

In Fig.\ref{fig: figure_elegans}, left, we observe that co-aging induced mortality is more pronounced in the early stages of life. As a significant portion of infected individuals are eliminated, an increase in intrinsic-damage induced mortality emerges where co-aging mortality is at its peak. This increase eventually plateaus. 

Note that our model not only generates non-monotonic cumulative hazard curves but can also disentangle the cause-specific hazard curves that underlie the ``bumps'' and ``kinks'' in the cumulative mortality curves.

%Last thing to emphasize is that increase of noise in all theoretical and experimental plots is due to the decrease of cohort size as we approach older ages.

{\bf Suppression in trees as a form of co-aging.}
\begin{figure}%[tbhp]
\centering
\includegraphics[width=\linewidth]{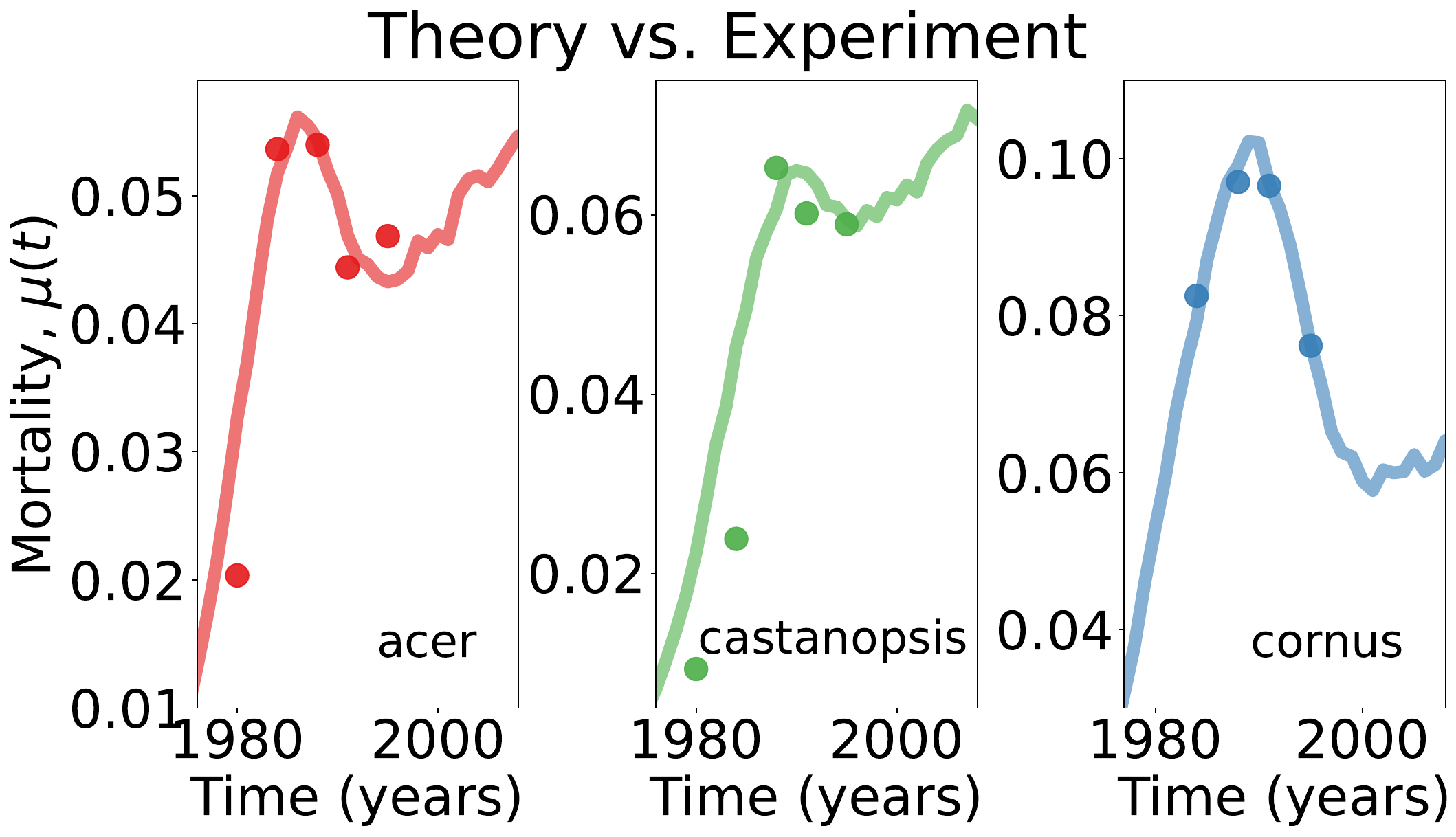}
\caption{\textbf{Co-aging of 3 tree species with their competitors (dots), fit to co-aging networks (curves).} The non-monotonic hazard curves of \textit{Acer macrophyllum} and \textit{Castanopsis chrysophylla} and \textit{Cornus nuttallii} \cite{Lutz_2006}, characterized by a decrease in mortality rates around the mid-stages of the observed period. Our model successfully replicates this trend by showing a similar decline during the intermediate stages of thinning. While trees are affected by the cumulative competition arising from the neighboring trees, our model simplifies these complex relationships by reducing all competitors to a single representative system. For model parameters, see supplemental material.}
\label{fig:tree fit}
\end{figure}
There are numerous factors contributing to tree mortality, ranging from fires, winds, drought, and predation \cite{Hood_2018, McDowell_2018, Bréda_2006, Sala_2010} to less apparent factors such as competition from other trees, pathogens and insects \cite{Peet_1987, Lutz_2006, Das_2016, Luo_2011, McDowell_2011, Anderegg_2015}. Some damaging factors occur uniformly throughout the lifespan of the tree, while others are age dependent \cite{Franklin_1987}. Our primary focus here will be on the ``thinning'' or so called ``stem-exclusion'' stage, where one of the most dominant mortality drivers is the competition for other trees for resources \cite{Lutz_2006, Franklin_1987, Peet_1987, McMurtrie_1981}. Suppression is a mortality factor that poses a lethal threat not only on its own but also by exacerbating other factors. This means that even when competition itself is not the direct cause of death, it can potentially intensify the adverse impacts on trees resulting from factors like climate change \cite{Ruiz-Benito_2013}.

In \cite{Lutz_2006}, 22 years of tree growth and mortality data was collected from early successional forests. The mortality curves they measured exhibited a non-monotonic pattern for all six tree species under investigation. This study provided empirical evidence that suppression stands as the most prevalent form of mortality during the initial phases of stand development.

In Figure \ref{fig:tree fit}, we compare our theoretical model to three of their data sets, fitting for \textit{A. macrophyllum}, \textit{C. chrysophylla} and \textit{C. nuttallii}. The mortality values reported in \cite{Lutz_2006} were defined as
$
    m(t_i) = 1-(S({t_i})/S({t_f}))^{1/ \Delta t}
$
where $m(t)$ is the mortality, $S(t)$ is the number of alive individuals and $\Delta t = t_f-t_i$ with $t_i$ and $t_f$ corresponding to time of initial and final measurement times, respectively. In our simulations each $\Delta t=1$, thus the formula reduces to our definition for mortality calculation $\mu(t) = [S(t)-S(t+1)]/S(t)$. For a discussion of definitions of mortality in this context, see \cite{Sheil_Burslem_Alder_1995}.

\begin{figure}%[tbhp]
\centering
\includegraphics[width=\linewidth]{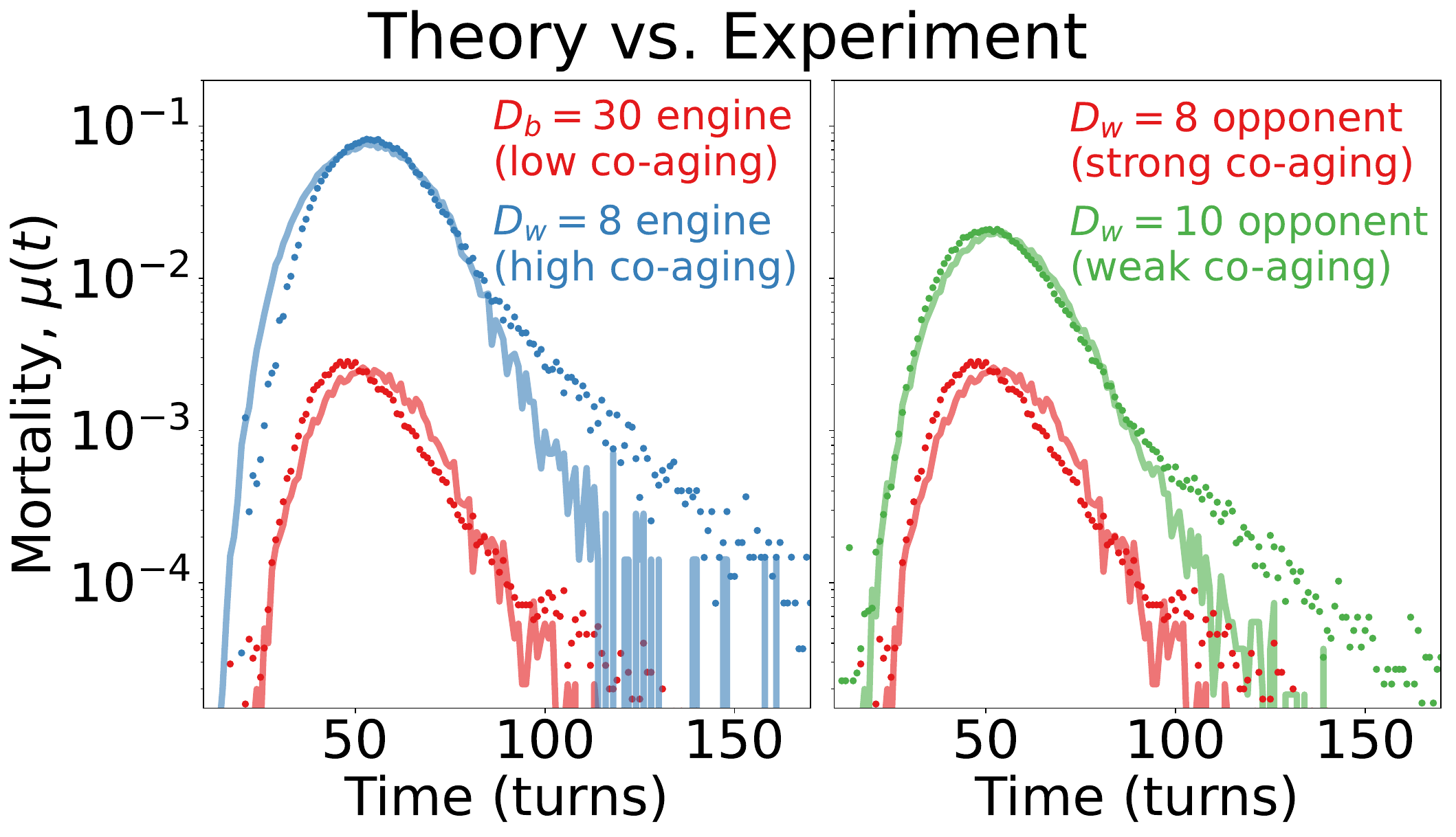}
\caption{\textbf{Co-aging of chess engines.} Left: We collected the failure statistics of two chess engines playing $400,000$ games against each other, where whites have thinking depth, $D_w = 8$ and the blacks, $D_b=30$. We achieved good fits to our model when we set $d=0$ (no damage source other than that of the antagonist), $f(0)=1$ (no prenatal damage), $r=0$ (no repair) which are reasonable assumptions for a chess game. Right: The black's hazard rate elevates, when its opponent white has a higher thinking depth. To fit the elevated hazard curve (green), we modified only the co-aging parameters in our simulations, such that the black-representing network receives more damage from the white and the white-representing network receives less damage from the black. The other parameters were kept constant (see supplementary material).}
\label{fig: chess_fit}
\end{figure}

{\bf Co-Aging in non-biological complex systems: Chess engines.} We expect to see the demographic signatures of co-aging whenever two complex systems with high interdependency damage one other, regardless of whether they are biological or not. Examples would include rivaling companies, armies, social organizations, sports teams, and countries.

Here, we test this claim by turning to a simple model system, chess, and leave the analysis of the other complex systems to future studies.

To obtain hazard curves in high resolution, we facilitate games between pairs of \emph{Stockfish} chess engines, and record the lifespan of players. One of the crucial parameters characterizing the strength of a chess engine is its thinking depth $D$, which quantifies the number of moves the engine can anticipate while calculating the best possible move. An engine with higher depth is able to inflict more damage per move, and receive less. Thus we associate the thinking depth with our co-aging parameters $C_{A,B}$. 

In chess, pieces organize to protect each other, forming dynamic networks. As opponents attack each others' network of defenses, they should co-age in a similar way to their biological analogs and our simulated interdependency networks. These games provide us with a highly controlled and tunable experimental environment where the relative strength of players can be adjusted, health status can be easily be monitored as a function of time, and population-scale measurements can be collected rapidly. 

However there are a number of subtle differences between chess engines and an organism. In chess, when one party loses, the game ends for both the winner and loser; whereas in biology, the winner continues to live, and better so, in the absence of an adversary. Thus, to analyze chess games and biological species on equal footing, we let the winning side continue to ``live'' after the game ends. We adopt this convention in our simulations too, i.e., when one network kills the other network, we let the winner continue to live instead of ending the simulations.

Second, our network model is not meant to describe the very particular end game mechanics of chess, where it is not loss of nodes but the immobility of a special node, the king, that defines death. Furthermore, towards the end, when there are only a few pieces left on the board, these no longer form an interdependent network. Instead the pieces fulfil their function in solitude, instead of part of a network.  For this reason, we proclaim a game lost for the party whose material strength first drops below a threshold (6 points), consistent with the condition of death for our simulated networks. The removal of this threshold condition introduces a boost in late-life tail in hazard curves, caused by persistent isolated loops; but the effect of removing the threshold is very similar in our theoretical model as well (see supplemental material).

Lastly, unlike biological organisms, chess armies do not experience any other damage apart from that inflicted by their opponents, $d_A=d_B=0$; they do not start with any ``prenatal'' damage, $f_A(0)=f_B(0)=1$; cannot repair themselves, $r_A=r_B=0$; and their sizes are precisely equal $N_A=N_B$. Moreover, if we were to change the thinking depth of a player, to update our theoretical fit we only allow ourselves to modify the co-aging damage parameters $C_{A,B}$. For example, if the thinking depth of $A$ is increased, we only allow $C_A$ (damage received by A) to be decreased and $C_B$ (damage inflicted by A) to be increased, keeping all other parameters the same.

The model fits reasonably well despite these severe parametric restrictions  (Fig.\ref{fig: chess_fit}). Here, the left panel displays the probability of losing the game as a function of turn number, as a chess engine of depth $D_w = 8$ (white) plays against one of depth $D_b=30$ (black). The right panel of Fig.\ref{fig: chess_fit} displays the mortality rate of black engines with $D_b=30$ against two different white engines with $D_w=8$ and $10$ using only two free parameters $C_A$ and $C_B$, due to our self-imposed restrictions.

\par

%Chess exhibits a fundamental distinction from the biological systems we have examined. In biological species, when one individual dies, the surviving entity continues to age due to the intrinsic damage accumulated throughout its lifetime. In contrast, a chess game concludes once one of the parties loses, and there is no inherent aging process. Therefore, in chess, intrinsic aging is not a factor, and the victorious side continues to exist indefinitely. Only damage considered in chess is the one that is inflicted by the other side. Another difference is the possibility of a draw in the game. To ensure consistency in our analysis, we filtered out games where it is concluded with a draw due to stalemate or insufficient material. 
\par
%Considering the nature of chess, when we compare our simulations to the outcomes observed in the game, we specifically focused on co-aging damage and did not consider intrinsic or prenatal damage. So, after one of the networks die, the other kept continue living indefinitely, similarly to chess. 

\begin{figure*}%[tbhp]
\centering
\includegraphics[width=1\linewidth]{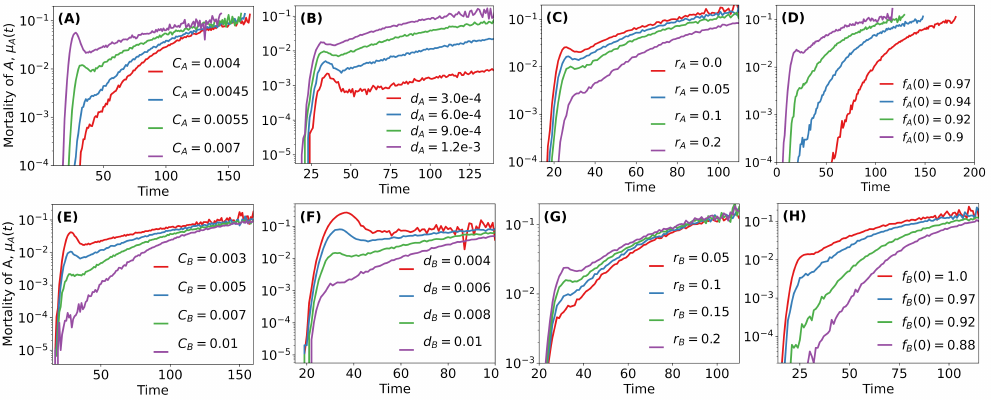}
\caption{{\bf The effect of system parameters on hazard curves.} (A) Increasing the co-aging constant $C_A$ results in the formation of a ``kink'' followed by a ``bump,'' which corresponds to a local maximum and increased probability of death in early life due to co-aging. (B) Increasing the intrinsic damage rate $d_A$ causes a shift in the hazard curve upwards. (C) As the repair rate, $r_A$, increases, both co-aging and intrinsic aging effects dissipate. (D) Decreasing prenatal damage, $f_A(0)$, leads to an increase in mortality and highlights the co-aging effects with the formation of a ``bump''. 
(E) Increasing the $C_B$ results in the dissipation of bump due to increased probability of death of $B$ in early life. (F) Increasing the intrinsic aging rate $d_B$ causes a shift in the hazard curve downwards. (G) As the repair rate, $r_B$ increases, both co-aging and intrinsic aging effects are emphasized. (H) Decreasing prenatal damage, $f_B(0)$ leads to an decrease in mortality and highlights the co-aging effects with the dissipation of the ``bump''. Details of the parameters for each of the curves provided in supplemental material.}
\label{fig: network_params}
\end{figure*}

{\bf A closer look at the effects of individual simulation parameters.} The top row of Fig.\ref{fig: network_params} (A-D), describes the influence of system parameters on the hazard curves of $A$. The co-aging constant emerges as a crucial determinant of the qualitative shape of the hazard curve, with increasing values resulting first in a ``kink'' followed by a ``bump'', as shown in panel (A). This effect can be attributed to $A$ becoming more adversely affected by $B$, causing a local maximum early on. Moreover, we observe the curve shifting to the left and reaching a plateau earlier as $C_A$ increases. In panel (B), we observe that the increase of the intrinsic aging rate $d_A$ of the network causes the hazard curve to shift upward, including the bump without changing its $x$ coordinate. This shift can be attributed to the increasing effect of intrinsic aging compared to co-aging with the increase of $d_A$. In contrast, increasing the repair rate of the networks $A$ diminishes the effect of both aging types. As shown in panel (C), the ``bump'' turns into a ``kink'' and eventually disappears completely, resulting in a reduced plateau. In panel (D), we examine the effect of prenatal damage, quantified by the initial number of functional nodes $f(t=0)$. An increase in prenatal damage highlights the co-aging effects on the system, leading to the appearance of a ``bump'' with emphasized co-aging for lower values of $f(t=0)$.

The bottom row of Fig.\ref{fig: network_params} (E-H) presents the influence of system parameters of type $B$ networks on the hazard curves of type $A$ networks. In panel (E), as the co-aging constant $C_B$ increases, type $B$ networks exhibit a diminished co-aging effect on type $A$ networks. Consequently, the previously observed bump in the hazard curve diminishes and eventually disappears as this parameter increases. Notably, unlike the case of changing $C_A$, the position of the bump remains unchanged. In panel (F), we observe that an increase in the intrinsic aging rate $d_B$ of network $B$ causes the bump to vanish without affecting the saturation value of mortality. On the other hand, increasing the repair rate of network $B$ intensifies the bump without altering the initial and saturation mortalities. Panel (G) demonstrates the transformation of the ``bump'' into a ``kink'', which eventually vanishes with decreasing repair rate of network $B$. Panel (H) shows that an increase in prenatal damage diminishes the co-aging effects on the type $A$ mortality. This shift results in postponing the initial appearance of mortality and shifting the curve to the right.

\section{Conclusion}
We introduced the concept of co-aging as the mechanism behind the anomalous features observed in the hazard curve of some organisms, which can be explained by neither evolutionary theories nor simpler reliability and interdependence network theories. We demonstrated however, that incorporating the notion of co-aging into the interdependence picture, can. %Through the development of our model, we were able to define two distinct types of death that a system can experience. One type is primarily caused by the system's intrinsic aging process, while the other with the failure induced by co-aging damage. By making this distinction, we were able to demonstrate the presence of these two types of hazard curves and discuss their relevance in biological systems.

To provide empirical support for our thesis, we fit 3 very different co-aging systems to our model. We were able to go beyond one fit per system, since our framework could make quantitative sense of experimental variables within individual datasets, such as treating the worms with antibiotics, cause-specific deaths, varying the thinking depth of chess engines, or varying the species of trees. Our framework was also able to go beyond fitting just cumulative hazard curves. Instead, we were able to fit independently measurable cause-specific components that add up to build the hazard curve. %Firstly, we investigated the aging patterns of \textit{C. elegans}, which exhibits qualitatively different hazard curves when co-aging with its pathogen \textit{E. coli}. The aging demographics of the worms, both in the presence and absence of the pathogen, is succesfully described by our theory. Lastly, we examined machine-vs-machine chess games under the lens of our co-aging theory and obtained good agreement.

The bumps and kinks observed in the hazard curves across diverse living and non-living complex systems is a potential   signal for co-aging. As such, we propose that aging demographics can be used as a kind of microscope with which ecological interactions can be probed. %By utilizing two different chess engines, we calculated the probability of losing for the weaker engine. We interpreted this probability of losing as a hazard curve, as the analogy of losing the game corresponds to the demise of the system. Remarkably, we observed a similar qualitative behavior in the hazard curves for the chess game. 

We end with a reminder that our model, like all models, is a simplified representation of a phenomenon, which in reality is extremely complex: we have not taken into account the time-dependent nature of interdependency networks (chess pieces move, immune systems adapt), the heterogeneity of nodes (not all cells/pieces have the same value and impact), the strategic (non-random) nature of attacks and defenses, and potentially many other interesting aspects of co-aging. Nevertheless, and despite these handicaps, we were able to unite within one theoretical framework, sickly worms, competing trees and chess battles.

\section*{ACKNOWLEDGMENTS}
We acknowledge the assistance of the Notre Dame Center for Research Computing staff and resources.

\bibliography{biblio}% Produces the bibliography via BibTeX.

\end{document}

\title{Supplemental Material}

\date{}
\maketitle

\begin{figure}[h]
\centering
\includegraphics[width=0.5\linewidth]{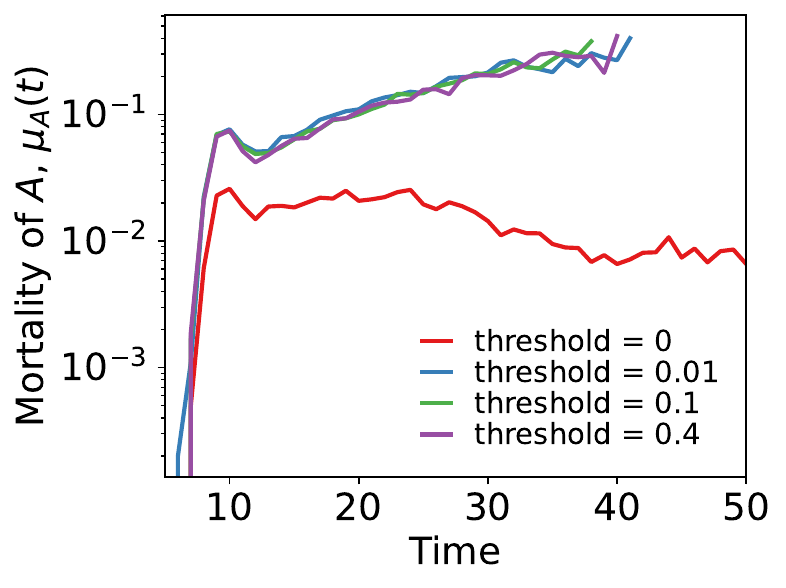}
\caption{\textbf{Change in mortality curves for different death thresholds.} When the threshold for defining the network as alive with the ratio of functional nodes is reduced to zero, we observe a distinctive shift in the shape of mortality curves. This change can be attributed to the formation of small cycles during the network's initial construction phase. What we predominantly observe is a network collapse occurring after approximately 50-60\% of the functional nodes have become non-functional. This results in a substantial reduction in the number of operational nodes, although they do not completely vanish. Importantly, we notice that when the nodes within these small cycles manage to persist, they exhibit a longer lifespan. Parameters used are same for the all four curves and given as $N_A = 3000$, $N_B=500$, $C_A=0.023$, $C_B=0.012$, $d_A=0.0037$, $d_b=0.025$, $r_A=0.0025$, $r_B=0.1$ and $f_A(0)=f_B(0)=1$ with the number of simulations $10,000$ for each curve.}
\label{appx: threshold change}
\end{figure}

{\bf Discussion of death threshold}. We proclaim a network to be dead if it's number of functional nodes drops below 1\%. A threshold value is occasionally since the network structure can include a small isolated cycle or clique that lingers on for some time even after the rest of the network has failed. However, such structures are extremely rare and  our plots remain unchanged for threshold values of 1\% to 40\% (Fig. S\ref{appx: threshold change}). This is because networks fail with a sudden collapse at the instant of death, loosing more than half of their nodes within one time step (check Figure 2 insets in paper). When this threshold value is set to zero however, even the small differences in network structure can cause highly differentiated mortality curves. This is due to a very small number of loops that are relatively disconnected from the rest of the network, and are not affected by the collapse of the whole. The larger a network, the less likely these persistent structures exist.

This phenomena reminds us of beetles being able to survive for many days after their heads are chopped off. 

As such, for small, decentralized networks, the distinction between life and death are not very clear cut, and a using a threshold to determine death time is practical.

When chess games are allowed to continue until the very end, some games extend to an extent where material damage ceases. Instead, a few pieces are chasing the king down for a long time, aiming to immobilize it.

The endgame in chess is qualitatively different than biological co-aging and our network model, since the condition for death is material (e.g. lack of pieces), but the immobility of a ``special'' node, the king. Furthermore, towards the end of the game, the piece density on the board reduces to a point where pieces no longer form an interdependent network, but roam freely, performing their function as disconnected individuals.
\begin{figure}[h]
\centering
\includegraphics[width=0.5\linewidth]{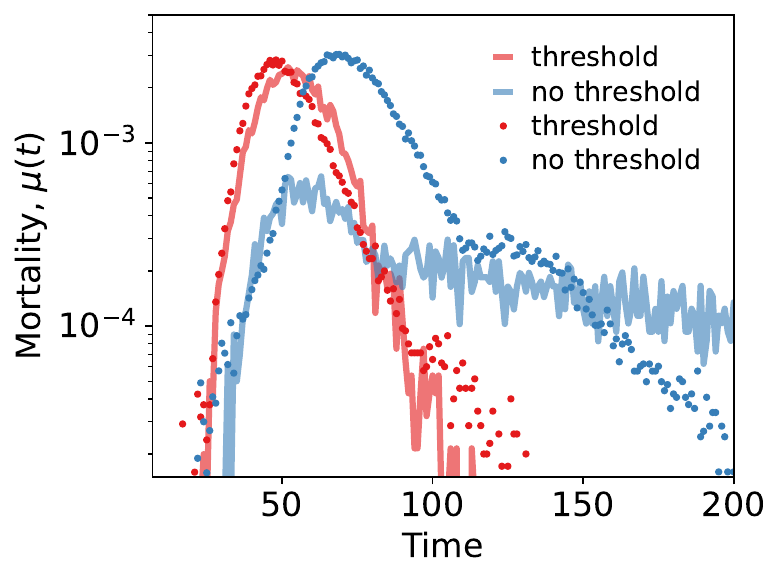}
\caption{\textbf{Change in chess and network mortality curves in the absence of threshold.}
Removing the material strength threshold results in an extended tail in the chess hazard curves (dots). This elongation is primarily attributed to prolonged games played between opponents with closely matched and low material scores, where neither is able to achieve checkmate for an extended period. In network simulations (lines), removing the vitality threshold of $0.1$ leads to a change, manifesting as persistent loops that endure for a prolonged duration in some networks. Chess curves show the probability of losing for the black side played by an engine with a depth of $30$ whereas its opponent white side engine has the depth $8$. For network simulations parameters are  $N_A = N_B = 350, C_A=3.3 \times 10^{-3}, C_B=5.5225 \times 10^{-3}$ with number of simulations and chess games equal to $100,000$.} 
\label{appx: thold_compare}
\end{figure}
For these reasons, we adopt a number of conventions that enable us to compare chess engines and organisms on equal footing, as discussed in the main article. One of these conventions is the loosing criteria: We proclaim a side lost not if their king is immobile, but if their material strength drops below a certain threshold. This convention is, in many ways, more similar to biology, and our network simulations. In our chess simulations, to assess  material strength, we assign conventional point values to chess pieces: Pawn=1, Knight=3, Bishop=3, Rook=5, and Queen=9. If, during the game, either side drops below a total material strength of 6 (e.g. three pawns and a knight) while the other is still above the threshold, we stop and conclude the game.

In this supplamentary section, we show that the qualitative behavior of hazard curves remain similar upon adopting this convention, except for a long tail at end life, presumably caused by a small number of pieces chasing around a king to immobilize it. Fig.S\ref{appx: thold_compare} shows the differences in hazard curves with and without the threshold for both chess games and network simulations. 

Interestingly, a comparable phenomenon occurs when we omit the 1\% threshold in our simulations as well, especially when the network is smaller. Similar to chess, this is caused in our simulations by long-lived loops that are relatively disconnected from and unaffected by the rest of the network. Since they can sustain independently from the rest, they do not get killed even when the rest of the system is dead, but instead must be targeted specifically (in a way, like the king). 

It is plausible that we might get better fitting simulation curves, if we designated one or more nodes in our networks as ``special'' and proclaimed the system dead upon loosing those nodes. The network could also re-organize to continue supporting these special nodes. However we do not explore such modifications to the model.

\begin{figure*}[h]
\centering
\includegraphics[width=\linewidth]{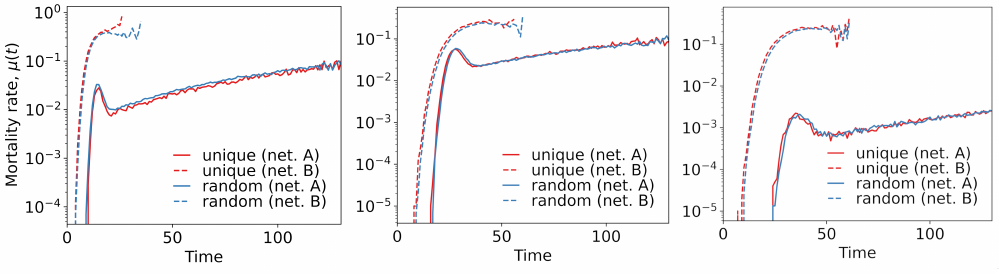}
\caption{\textbf{Comparison of mortality rates of among populations formed with identical networks and randomized networks.}
Three different figures correspond to different set of parameters. For the population formed of randomized networks each network has the same number of nodes but their connections are established differently from one another. For the population with only one unique network we use the same network for each aging process. In all three figures we see no significant difference between mortality curves of randomized population and unique network population. In a figure red and blue curves correspond to mortality of network A and network B, respectively. Parameters used are given as: (Left) $N_A = 2000$, $N_B=350$, $C_A=0.015$, $C_B=0.0$, $d_A=0.0011$, $d_b=0.02$, $r_A=r_B=0$, $f_A(0)=f_B(0)=1$, (Middle) $N_A = 2500$, $N_B=500$, $C_A=0.007$, $C_B=0.0$, $d_A=0.001$, $d_b=0.008$, $r_A=r_B=0.0$ $f_A(0)=f_B(0)=1$ and (Right) $N_A = 2500$, $N_B=500$, $C_A=0.0055$, $C_B=0.0$, $d_A=0.0003$, $d_b=0.008$, $r_A=r_B=0.0$  $f_A(0)=f_B(0)=1$ with the number of simulations $100,000$ for all.} 
\label{appx: uniq_vs_random}
\end{figure*}

\textbf{Effect of differences in network topology.}
While carrying out the simulations to calculate mortality throughout the paper we chose to age identical networks over and over again by acknowledging the fact that any two biological species should not differ in terms of the connections between its constituents. Here we are testing the what happens when any two random individual in the population has different connections versus when we use only one unique network for the whole population. Fig. S\ref{appx: uniq_vs_random} shows that we don't see any significant change for any of the three simulations. Each inset in the figure corresponds to a different simulation with different parameters where blue lines correspond to the mortality when networks are randomized and red lines show when only one unique network is used. 

\bf{Parameter list for Figures.}
%\textbf{Figure \ref{fig: network_params} parameters:} $N_A=2500$, $N_B=500$ and number of simulations $S=100,000$ for all. \textbf{(A)} $C_B=0.0$, $d_A=0.001$, $d_B=0.008$, $r_A=r_B=0$, $f_A(0)=f_B(0) = 1.0$. \textbf{(B)} $C_A=0.0055$, $C_B=0.0$, $d_B=0.008$, $r_A=r_B=0$, $f_A(0)=f_B(0) = 1.0$. \textbf{(C)} $C_A=0.007$, $C_B=0.0$, $d_A = 0.0015$, $d_B=0.01$, $r_B=0$, $f_A(0)=f_B(0) = 1.0$. \textbf{(D)} $C_A=0.003$, $C_B=0.0$, $d_A=0.0009$, $d_B=0.012$, $r_A=r_B=0$, $f_B(0) = 1.0$. \textbf{(E)} $C_A=0.007$, $d_A=0.001$, $d_B=0.006$, $r_A=r_B = 0.0$, $r_A=r_B=0$, $f_A(0)=f_B(0) = 1.0$. \textbf{(F)} $C_A=0.0055$, $C_B=0.0$, $d_A=0.0011$, $r_A=r_B = 0.0$, $r_A=r_B=0$, $f_A(0)=f_B(0) = 1.0$ \textbf{(G)} $C_A=0.0055$, $C_B = 0.0$ $d_A=0.0015$, $d_B=0.01$, $r_A=0.03$, $f_A(0)=f_B(0) = 1.0$ \textbf{(H)} $C_A=0.005$, $C_B =0.0$, $d_A=0.0014$, $d_B=0.01$, $r_A, r_B=0.0$, $f_A(0)=0.97$ %

\textbf{Fig.2:} For left curve, $C_A=0.015$, $C_B=0.0$, $d_A=0.0011$, $d_B=0.02$, $r_A=r_B=0$, $f_A(0)=f_B(0) = 1.0$. For the right curve only $C_A=0.0$ is different and the rest of the parameters are same. $N_A=2000$, $N_B=350$ and number of simulations $S=100,000$ for both

\textbf{Fig.3:}
$C_A = 0.023$, $C_B=0.012$, $d_A=0.0037$, $d_B=0.025$, $r_A=0.0025$, $r_B=0.1$, $f_A(0)=f_B(0)=1.0$, $N_A=3000$, $N_B=500$ and the number of simulations $S=10,000$.

\textbf{Fig. 4:} From left to right (Number of simulations $S=200,000$ for all): 

\textbf{    Acer}: $C_A=0.0035$, $C_B=0.0005$, $d_A=0.0017$, $d_B=0.01$, $r_A=0.07$, $r_B=0$, $f_A(0)=0.9$, $f_B(0) = 1.0$, $N_A=N_B=700$, 

\textbf{Castanopsis}: $C_A=0.00355$, $C_B=0.0014$, $d_A=0.0024$, $d_B=0.008$, $r_A=0.07$, $r_B=0$, $f_A(0)=0.905$, $f_B(0) = 1.0$, $N_A=N_B=700$, 

\textbf{Cornus}: $C_A=0.0038$, $C_B=0.0$, $d_A=0.0015$, $d_B=0.008$, $r_A=0.0$, $r_B=0$, $f_A(0)=0.9$, $f_B(0) = 1.0$, $N_A=N_B=700$.

\textbf{Fig.5 (left):} $C_A=3.6 \times 10^{-3}$, $C_B=5.55 \times 10^{-3}$, $N_A = N_B = 350$, rest of the parameters are zero and number of simulations $S=100,000$. 

\textbf{Fig.5 (right):} $C_A=4.5 \times 10^{-3}$, $C_B=4.655 \times 10^{-3}$, $N_A = N_B = 350$, rest of the parameters are zero and number of simulations $S=100,000$. 

\textbf{Fig.6} parameters given below in the table.

\begin{table*}[h]
\centering
\begin{tabular}{l*{10}{c}r}
Panel             & $C_A$ & $C_B$ & $d_A$ & $d_B$ & $r_A$  & $r_B$ & $f_A(0)$ & $f_B(0)$ \\
\hline
\textbf{(A)} & $-$ & $0.0$ & $0.001$ & $0.008$ & $0.0$ & $0.0$ & $1.0$ & $1.0$  \\
\textbf{(B)}            & $0.0055$ & $0.0$ & $-$ & $0.008$ & $0.0$ & $0.0$ & $1.0$ & $1.0$  \\
\textbf{(C)}           &$0.007$ & $0.0$ & $0.0015$ & $0.01$ & $-$ & $0.0$ & $1.0$ & $1.0$ \\
\textbf{(D)}     & $0.003$ & $0.0$ & $0.0009$ & $0.012$ & $0.0$ & $0.0$ & $-$ & $1.0$  \\
\textbf{(E)} & $0.007$ & $-$ & $0.001$ & $0.006$ & $0.0$ & $0.0$ & $1.0$ & $1.0$  \\
\textbf{(F)}            & $0.0055$ & $0.0$ & $0.0011$ & $-$ & $0.0$ & $0.0$ & $1.0$ & $1.0$  \\
\textbf{(G)}           & $0.0055$ & $0.0$ & $0.0015$ & $0.01$ & $0.03$ & $-$ & $1.0$ & $1.0$  \\
\textbf{(H)}     & $0.005$ & $0.0$ & $0.0014$ & $0.01$ & $0.0$ & $0.0$ & $0.97$ & $-$ %  \\
\end{tabular}
\end{table*}

